# Composite based magnetoelectric scaled devices with large output voltages

Emma Van Meirvenne,[1,2] Aude Brinkmann-Hornbogen,[2] Bart Sorée,[1,2] Christoph Adelmann,[1] and Florin Ciubotaru[1]

[1]*Imec, 3001 Leuven, Belgium*

[2]*KU Leuven, Departement Elektrotechniek, 3001 Leuven, Belgium*

## Abstract

In this work, we investigate the differential voltage generation arising from the direct magnetoelectric (ME) effect in nanoscale composite devices upon magnetization rotation from the magnetic ground state to an out-of-plane (OOP) configuration. These composite devices comprise a magnetostrictive ferromagnetic layer and a piezoelectric layer, mechanically coupled through strain. Using a finite element method (FEM) model, developed in COMSOL Multiphysics, we provide a comprehensive analysis of strain transfer mechanisms and resulting voltage generations. Here, the influence of dimensional and material parameters on the device performance is systematically examined. Our results indicate the presence of two distinct strain transfer mechanisms at scaled dimensions, where the device aspect ratio and the magnetic state both determine the dominant mechanism influencing the strain transfer to the piezoelectric layer. Moreover, we observed that the influence of surface clamping diminished as the pillar area was reduced. We also saw that the strain transfer to the piezoelectric layer can be enhanced by using stiffer electrodes or clamping layers. Lastly, we concluded that magnetostrictive materials with large magnetoelastic coupling constants or large Poisson ratios may strongly increase the output voltage at small dimensions. This study provides insight in the dimension and material selection when designing scaled ME pillars, with the aim of generating large output voltages. We showed that output voltages exceeding 200 mV can be achieved in scaled devices, underscoring the potential of these structures for integration into microelectronic applications.

## I. INTRODUCTION

Composite magnetoelectric devices, consisting of a piezoelectric and a magnetostricitve layer, have gained interest due to their possibility for strong magnetoelectric coupling at room temperature [1–5]. In these composite systems, the magnetoelectric coupling is mediated through strain, enabling the converse magnetoelectric (ME) effect, wherein strain generated in the piezoelectric layer is transferred to the magnetostrictive layer, leading to a rotation of its magnetization [6–9]. Magnetization can also be reoriented in single-phase multiferroics [10, 11], which intrinsically exhibit ferroelectric and (anti-)ferromagnetic order [2, 12–15]. However, only a limited number of such materials remain multiferroic at room temperature [16]. The most studied multiferroic material is *BiFeO$_3$*, with has weak ME coupling [17]. In contrast, magnetoelectric composite systems can achieve substantially stronger coupling, as the magnetostrictive and piezoelectric constituents can be individually engineered to optimize their respective properties and thereby enhance the overall magnetoelectric response [18]. Furthermore, these composite systems also exhibit the converse effect, which is better known as the direct ME effect, where the polarization is controlled by an external magnetic field. [19–23]. Recently, the direct ME effect is shown experimentally in large-area ME composites, showing voltage generation of approximately 70 mV when rotating the magnetization from an in-plane (IP) to an out-of-plane (OOP) state [24]. The direct ME effect is also observed in multiferroics, but their ME coupling coeffients are low [25]. Because ME composites exhibit both direct and converse ME effect at room temperature and their ME coupling coefficient can be large by tuning the separate composite layers, they are promising materials for low-power spintronic memory, logic devices, MEMS and sensors [10, 11, 26–31].

For the direct ME effect to be useful for microelectronic applications, the ME coupling coefficients need to be enhanced, so large voltages can be achieved. Since the ME composite coupling is strain-mediated, an efficient strain transfer from the magnetostrictive to piezoelectric layer is required [10, 11, 26–30, 32, 33]. Regarding this strain transfer, one major challenge remains to limit substrate clamping, since it suppresses the IP strain transfer from the magnetostrictive to the piezoelectric layer [34]. Many solutions have been proposed to overcome this clamping [6, 35–43]. However, one other way to overcome substrate and surface clamping while retaining planar device architectures, which is not explored much yet, is geometric scaling. It is shown in finite-element method (FEM) simulations that when ME pillars are miniaturized to the submicron regime, edge relaxation can locally relieve the substrate constraint, allowing for more efficient strain-transfer [24, 44]. Moreover, the small ME pillar areas are promising, especially for nanoelectronics applications. Recently, an experimental demonstration of the quasi-static direct ME effect was shown in nanoscale ME composite device arrays on a rigid ME platform [45]. The differential voltage, which is taken as the difference in voltage obtained for an in-plane (IP) and out-of-plane (OOP) magnetization state, was measured to be in the order of several mVs. Despite this being an output voltage that is much larger than what is typically being obtained by the inverse Spin-Hall effect [11, 46], larger voltages are needed to be relevant for microelectronic applica-tions. Therefore, an in-depth study on the strain transfer, clamping mechanisms and voltage generation by the direct ME in scaled composite devices is required to enhance achievable output voltages.

This study investigates the strain transfer and voltage generation by the direct ME effect in nanoscale composite pillar devices on a rigid Si substrate upon rotating the magnetization from its magnetic ground state to an OOP state via FEM simulations. A FEM model is built in COMSOL Multiphysics that allows for multi-physics field coupling, and which is validated via experiments [24, 45]. Using this calibrated model, insights on the strain transfer mechanisms and clamping effects in nanoscaled devices are provided by varying device dimensions and materials. This study shows that voltages up to 100 mV can be achieved, even in scaled pillars. These findings aim to provide practical design guidelines for optimizing nanoscale ME devices to achieve high output voltages, relevant for sensors or applications in microelectronic circuits where energy efficient reading schemes are required.

## II. FEM MODEL OF ME DEVICE

The proposed device is consisting of a ferromagnetic magnetostrictive and a piezoelectric layer, with top and bottom metallic electrode, all shaped in a form of circular pillars. The device is surrounded by a soft material spin-on carbon (SOC), similarly to the experimental devices [45], and it is placed on a 3 μm thick $SiO_2$ layer to account for the substrate effects, such as substrate clamping. A schematic of the device is illustrated in Fig. 1. The magnetoelastic coupling is implemented by the body force acting on the magnetic volume and the mechanical boundary condition at the magnet's surface are given by [47–49]

$$f_{mec} = \begin{pmatrix} 2b_1 m_x \frac{\partial m_x}{\partial x} + b_2 \left( \frac{\partial m_x}{\partial y} m_y + \frac{\partial m_y}{\partial y} m_x + \frac{\partial m_z}{\partial z} m_x + \frac{\partial m_x}{\partial z} m_z \right) \\ 2b_1 m_y \frac{\partial m_y}{\partial y} + b_2 \left( \frac{\partial m_x}{\partial x} m_y + \frac{\partial m_y}{\partial x} m_x + \frac{\partial m_y}{\partial z} m_z + \frac{\partial m_z}{\partial z} m_y \right) \\ 2b_1 m_z \frac{\partial m_z}{\partial z} + b_2 \left( \frac{\partial m_y}{\partial y} m_z + \frac{\partial m_z}{\partial y} m_y + \frac{\partial m_z}{\partial x} m_x + \frac{\partial m_x}{\partial x} m_z \right) \end{pmatrix} \quad (1)$$

$$f_{mec} = -n \begin{pmatrix} b_1 m_x^2 & b_2 m_x m_y & b_2 m_x m_z \\ b_2 m_y m_x & b_1 m_y^2 & b_2 m_y m_z \\ b_2 m_z m_x & b_2 m_z m_y & b_1 m_z^2 \end{pmatrix} \quad (2)$$

where $b_1$ and $b_2$ are the magnetoelastic coupling coefficients, $m_x = M_x/M_S$, $m_y = M_y/M_S$ and $m_z = M_z/M_S$ are the normalized magnetization components, $M_S$ is the saturation magnetization and $n$ is the norm vector on the boundary. The magnetization ground state is calculated using the Landau-Lifshitz-Gilbert (LLG) equation utilizing a micromagnetics module developed in COMSOL [50, 51]. The obtained magnetization states are tested against MuMax3 micromagnetic solver [52]. For all studied structures, we have obtained similar distributions for the magnetization ground states in both MuMax3 and COMSOL, validating the magnetization distribution obtained by the COMSOL models. To implement magnetoelastic coupling by magnetostriction, the Micromagnetic module is coupled to the Solid Mechanics module by implementing Eqs. (1) and (2) [47, 49]. The Solid Mechanics module is eventually coupled to the Electrostatics module by the constitutive piezoelectric equations to account for piezoelectric coupling.

The material properties for the magnetostrictive layer and electrodes used in this work are summarized in Table I, and are assumed to be isotropic. Hence, the magnetoelastic coupling coefficient $B = b_1 = b_2$ [53]. ScAlN is chosen as piezoelectric material because of its relatively low value of electric permittivity, which is ideal for large voltage generation, according to the constitutive piezoelectric equations [54]. The following material parameters for ScAlN (assumed to be a hexagonal crystal) were considered in simulations: density $\rho$ = 3500 kg/m$^3$, isotropic permittivity $\epsilon$ = 16, stiffness constants $C_{11} = C_{22}$ = 280 GPa, $C_{12}$ = 135 GPa, $C_{13} = C_{23}$ = 120 GPa, $C_{33}$ = 160 GPa, $C_{44} = C_{55}$ = 100 GPa, $C_{66} = (C_{11} - C_{12})/2$, and piezoelectric stress constants of $e_{15} = e_{24}$ = -0.35 C/m$^2$, $e_{31} = e_{32}$ = -0.55 C/m$^2$ and $e_{33}$ = 2.7 C/m$^2$ [24, 55].

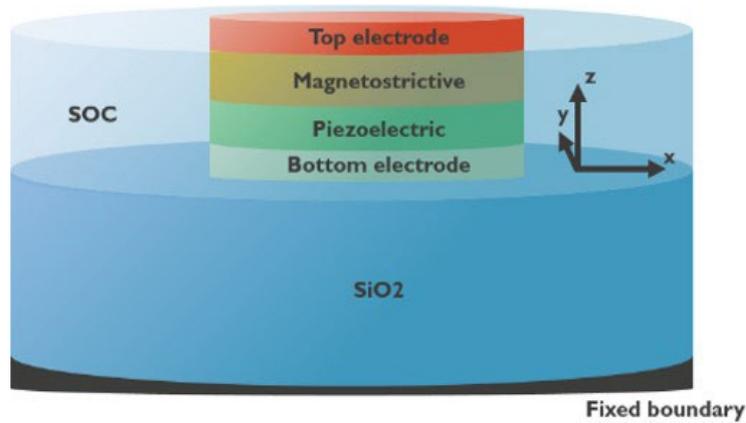

*Figure 1 Schematic of a ME device, positioned on a SiO2 substrate and surrounded by a soft material (spin-on carbon, SOC). Devices with pillar diameters ranging from 100 nm to 2000 nm, various electrode materials (Au, Ru), and different thicknesses and material choices for both the piezoelectric and magnetic layers (e.g., Ni, FeGa, Terfenol-D) were studied.*

| Material | $\rho$ (g/cm$^3$) | $E$ (GPa) | $v$ (1) | $B$ (MJ/m$^3$) | $M_S$ (kA/m) | $A_{ex}$ (pJ/m) |
|---|---|---|---|---|---|---|
| Ni [24] | 8.9 | 200 | 0.31 | 10 [56] | 480 [7] | 15 [57] |
| Fe$_{0.75}$Ga$_{0.25}$ | 7.8 [58] | 27 [59, 60] | 0.47 [60] | -20 [61] | 980 [61] | 13 [62] |
| Terfenol-D [53, 63] | 9.21 | 80 | 0.30 | -111 | 700 [64] [62] | 9 [64] |
| Au | 19.3 [24] | 70 [65] | 0.44 [65] | – | – | – |
| Ru | 12.37 [66] | 447 [67] | 0.25 [66] | – | – | – |
| SOC [18] | 2.22 | 14 | 0.25 | – | – | – |
| *SiO$_2$* | 2.2 | 70 | 0.17 | – | – | – |

TABLE I. Material parameters for Ni, Fe$_{0.75}$Ga$_{0.25}$ (FeGa), Terfenol-D, Au, Ru, SOC and SiO$_2$: mass density $\rho$, Young's modulus $E$, Poisson ratio $v$, magnetoelastic coupling constant $B$, satu-ration magnetization $M_S$, and Exchange constant $A_{ex}$. The material parameters for *SiO$_2$* were sourced from the COMSOL library.

## III. SIMULATION RESULTS

A 90° magnetization rotation is required to generate a maximal difference in magnetostrictive strain, whereas 180° rotations do not result in different strain distributions [53]. Therefore, the strain and voltage differences are calculated for magnetic ground states and OOP states.

As suggested by Eq. (1) and (2), the magnetization distribution strongly influences the generated magnetostrictive strain distribution in the magnet. Figure 2 demonstrates the difference in strain distribution for different magnetization states. For the case of a uniform in-plane (IP) magnetization state (Fig. 2 (a)), the longitudinal strain $\varepsilon_{zz}$ that is generated in the magnet is tensile, while the axial strain $\varepsilon_{xx}$ is compressive. Conversely, for the OOP magnetization case, $\varepsilon_{zz}$ in the magnet is compressive, while $\varepsilon_{xx}$ is tensile (Fig. 2 (c)). For both magnetization states, shear strain arises at the pillar edges, at the interface between the magnet and piezoelectric layer (see App. A Fig. 8( a) and App. C Fig. 10 (b)). This arising of shear strain is caused by a mismatch in displacement between the magnet and the piezoelectric layer due to axial loading of the magnet. These shear strains cause normal strain to arise in the piezoelectric layer, as seen in App. A Fig. 8 (a). Note that, for both IP and OOP magnetization states in Fig. 2, the strain at the edges in the piezoelectric layer changes to compressive/tensile strain with respect to the tensile/compressive strain in the pillar centre. This change is also what we refer to as bending, which decreases the average strain present in the piezoelectric layer. For the vortex magnetization state (Fig. 2 (b)), a strong tensile strain $\varepsilon_{zz}$ and a strong compressive strain $\varepsilon_{xx}$ arise at the vortex core in the magnet. In this case, large shear strain arises over the entire magnet-piezoelectric interface, due to the non-uniformity of the magnetization distribution (see App. C Fig. 10 (a)). This results in a non-uniform strain distribution in the piezoelectric layer center, as seen in Fig. 2 (b). For all magnetization states, the presence of shear strain at the magnet-piezoelectric interface is caused by a mismatch in displacement between the magnet and the piezoelectric layer due to axial loading of the magnet. To counteract these shear strains, axial strains arise in the piezoelectric layer. Therefore, the presence of shear strain at the interface allows for the transfer of the axial strain components $\varepsilon_{XX}$ and $\varepsilon_{YY}$ from the magnet to the piezoelectric layer [68–70]. This is explained in more detail in App. A. On the other hand, the longitudinal strain component, $\varepsilon_{zz}$, is indirectly transferred by shear strain, after being converted to axial strains through the Poisson effect. The Poisson ratio in isotropic materials is given as $v = -\varepsilon_{XX}/\varepsilon_{ZZ}$ [71].

The output voltage is generated by the piezoelectric effect and is defined by all normal and shear strain components $\varepsilon_{kl}$ present in the piezoelectric layer via the piezoelectric stress coefficients $e_{kij}$. This is illustrated in the stress-charge form of the constitutive equations [54]: $\sigma_{ij} = c^E_{ijkl}\varepsilon_{kl} - e_{kij}E_k$ and $D_i = e_{ikl}\varepsilon_{kl} + \epsilon_0\epsilon^\varepsilon_{r,ik}E_k$. In these equations, $\sigma_{ij}$ is the stress, $c^E_{ijkl}$ is the elasticity tensor at constant electric field, $D_i$ is the displacement field, $\varepsilon_{kl}$ is the strain, $\epsilon_0$ is the vacuum permittivity, $\epsilon^\varepsilon_{r,ik}$ is the vacuum permittivity at constant strain and $E_k$ is the electric field. Despite the different normal strain components being similar in magnitude, the normal strain component $\varepsilon_{ZZ}$ is the dominant term in defining the electric field due to the larger value of $e_{33}$ with respect to the other piezoelectric stress coefficients. Therefore, we have chosen for simplicity to display only $\varepsilon_{zz}$ in Fig. 2. Note that the strain being compressive or tensile does not depend on the magnetization polarity, since 180° magnetization rotations result in equal magnetostricitve strain generation [53].

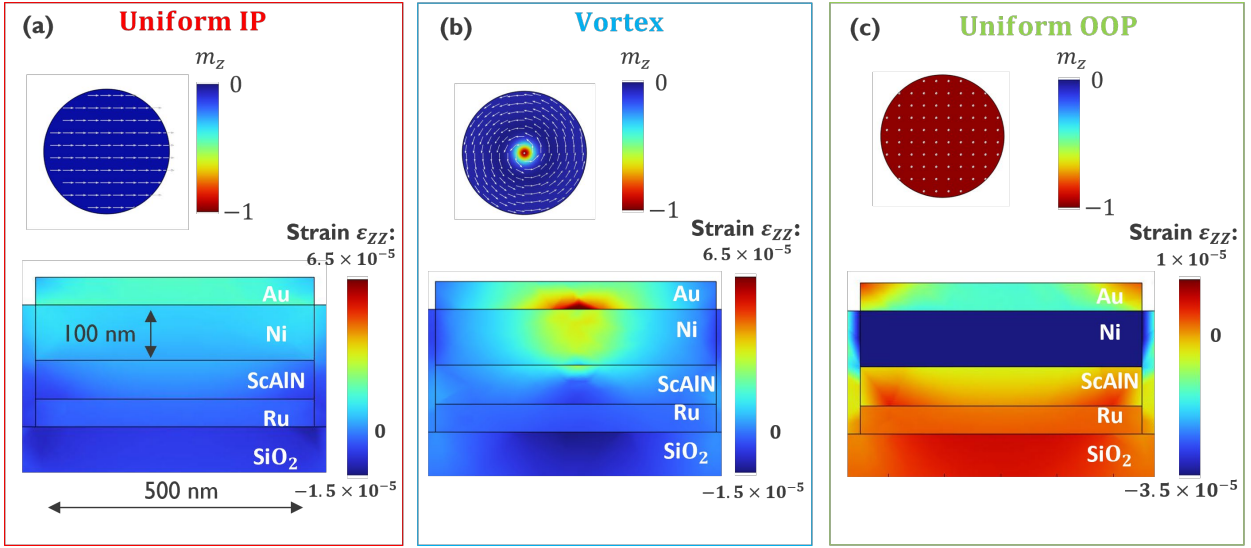

*Figure 2 Magnetization states with the $m_z$-component color coded (upper panels) and the normal strain component $\varepsilon_{zz}$ (bottom panels) in a 500 nm diameter device and 100 nm thick Ni magnet in different magnetic states: (a) uniform in-plane (IP) state , (b) vortex ground state, and (c) uniform out-of-plane (OOP) state, respectively.*

In a first step, we investigated the dependence of the strain transfer and the corresponding generated voltage on the pillar area. The results are displayed in Fig. 3. The differential voltage in Fig. 3 (a) is defined as the voltage difference between the out-of-plane (OOP) magnetization state and either the vortex state or a uniform in-plane (IP) magnetization state. It should be noted that, for the investigated pillar diameters, the vortex state is the natural magnetic ground state, whereas the IP or OOP states can be stabilized by applying bias magnetic fields along the corresponding directions. Figures 3 (b) and (c) illustrate the average longitudinal strain component, $\varepsilon_{zz}$, in the piezoelectric layer and the electric potential generation for the different magnetization states, respectively. It can be observed that the vortex state and the uniform IP state lead to a very similar voltage generation (Fig. 3 (c)), although the strain distribution is very different (see Fig. 2 (a) and (b)). Yet, in average the induced strain is very similar for the two magnetic states, as shown in Fig. 3 (b). Hence, the uniform IP and vortex magnetic states generate comparable electric potentials. Note that the main strain transfer mechanism for the IP and vortex magnetization state is shear strain transfer. At small diameters (*i.e.*, at reduced aspect ratios), the electric potential for the vortex and uniform IP state is small. This is because the axial strain that is transferred to the piezoelectric layer quickly decays to zero towards the piezoelectric layer depth, as seen in Fig. 3 (d) (see also App. A Fig. 8 (b)). Increasing the pillar diameter leads to a higher electric potential for both IP and vortex magnetization states, owing to the enhanced normal strain generated in the piezoelectric layer. Although the average magnetostrictive strain in the magnet decreases with increasing pillar diameter, due to increasing surface clamping effects, the strain transfer efficiency increases strongly (see App. B Fig. 9 (a) and (c)), because the normal strain penetrates deeper in the piezoelectric layer (App. A Fig. 8 (b)). This results in an increasing average normal strain in the piezoelectric layer, as seen in Fig. 3 (b). Around pillar diameters of 500 nm, the normal strain in the piezoelectric layer is slightly larger for vortex states, because the highly non-uniform vortex core generates large strains in the magnet center, leading to increased strain transfer to the piezoelectric layer (see Figs. 2 (b) and (c)). For larger pillar diameters, the influence of the non-uniform vortex core decreases (compare Figs. 3 (e) and (f)), and similar average strain is transferred for both magnetization states. At these larger pillar diameters, the strain present in

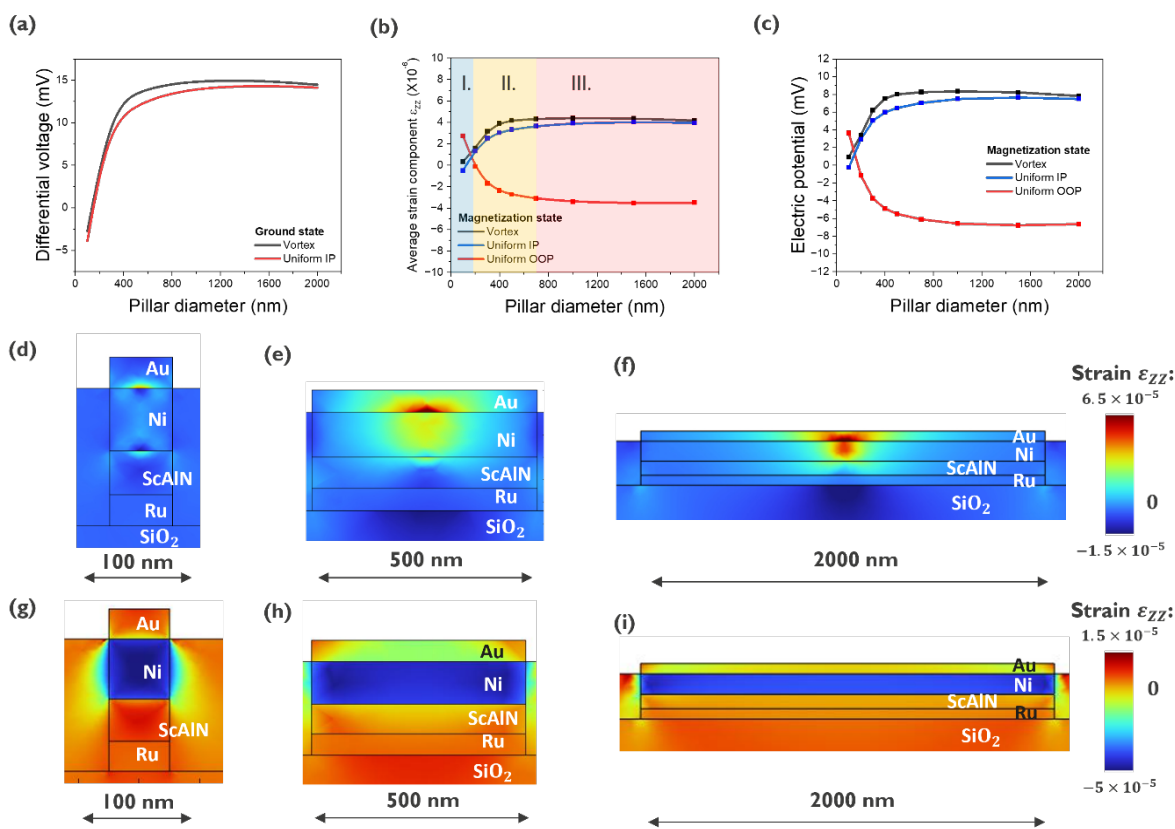

*Figure 3 (a) Differential voltage when the magnetization is switched from a uniform IP (red) or vortex ground state (black) to an uniform OOP state. (b) Average strain component $\varepsilon_{zz}$ in the piezoelectric layer for different magnetization states: vortex state (black), uniform IP (blue) and uniform OOP (red). Three different regimes are indicated: I - direct elongation of the piezoelectric layer by the OOP-compressing magnet yields tensile $\varepsilon_{zz}$; II - increasing device diameter enhances shear-induced normal strain, counteracting magnet-driven elongation and producing near-zero or compressive average strain; and III - shear-dominated strain transfer and surface clamping lead to a saturated compressive response. (c) Electrical potential at the top electrode for a vortex state (black), uniform IP (red curve) and uniform OOP (blue curve) state. The strain component $\varepsilon_{zz}$ for a vortex magnetization state for a diameter of (d) 100 nm, (e) 500 nm and (f) 2 µm, $\varepsilon_{zz}$ for an OOP magnetization state for a diameter of (g) 100 nm, (h) 500 nm and (i) 2 µm. The thicknesses of the magnet, piezoelectric layer and electrodes are set to 100 nm, 70 nm and 50 nm, respectively*

the piezoelectric layer saturates, due to increasing surface clamping of the magnet and enhanced strain transfer efficiency (see App. B Fig. 9 (a) and (c)), hence the generated electric potential saturates for increasing pillar diameters.

For OOP magnetization states and small pillar diameters, the average longitudinal strain $\varepsilon_{zz}$ in the piezoelectric layer changes from being compressive to being tensile (regime I in Fig. 3 (b)). This is because the magnet is compressing in z-direction, hereby directly elongating the piezoelectric layer in this direction, generating a tensile strain $\varepsilon_{zz}$ (see Fig. 3 (g)). When the pillar diameter is increased (and thus its aspect ratio), the compressing magnet becomes less effective at directly elongating the piezoelectric layer, because the normal strain generated by the shear components (as described in App. A) begins to counteract this elongation. These effects are compensating each other for pillar diameters around 200 nm resulting in a near zero average strain in the piezoelectric layer. Above this critical dimension, the piezoelectric layer starts to compress in the z-direction (negative average normal strain) as can be seen for Region II in Fig. 3 (b). Region III corresponds to pillar diameters for which the strain is generated entirely by shear strain at the edges and for which a saturation state is achieved due to surface clamping effects.

Diving deeper into geometrical effects on the pillar performance, we have investigated the impact of the magnet thickness on the differential voltage generation. Figure 4 (a) shows the average strain, $\varepsilon_{zz}$, in the piezoelectric layer as a function of pillar diameter for two magnet thicknesses: 100 nm and 200 nm. The corresponding differential voltage generation is shown in Fig. 4 (b). It can be observed that for pillar diameters around 200 nm and below, there is nearly no difference between the average strain for the ground state and the uniform OOP magnetization. This is because for the increased magnet thickness and reduced aspect ratio (diameter vs thickness) the shape anisotropy is reduced and, in the absence of the external field, the magnetization naturally aligns more in the z-direction. Figure 4 (c) shows the z-component of magnetization for pillars with 100 nm diameter and magnet layer thickness of 100 nm and 200 nm, respectively. It can be seen that for an aspect ratio of 0.5 the magnetization is close to a uniform OOP configuration, whereas for an aspect ratio of 1 the vortex state is clearly observed. For pillar diameters above 500 nm, the 200 nm thick magnet induces larger strains in average. This arises because surface-clamping effects are less pronounced in thicker magnetic layers. When a uniform in-plane magnetization state is enforced (e.g., by an external magnetic field or intrinsic in-plane anisotropy), the difference in the average strain within the piezoelectric layer between the IP state and its OOP state increases significantly, thereby strongly enhancing the differential voltage for a magnet thickness of 200 nm. Consequently, for very small pillar diameters (i.e., 100 nm and below), the differential voltage in Fig. 4 (b) would be larger in absolute value for a 200 nm-thick magnet compared to a 100 nm-thick magnet, provided that a vortex or IP magnetization state is ensured as the magnetic ground state.

For the OOP magnetization state, the three different regimes discussed in Fig. 3(b) can also be identified for the 200 nm-thick magnet. In regime I, the piezoelectric layer elongates in the z-direction due to compression of the magnet along this direction. When thicker magnets are used, a larger tensile strain $\varepsilon_{zz}$ is induced in the piezoelectric layer. As explained for Fig. 3 (b), bending of the piezoelectric layer occurs in regime II. This bending becomes more pronounced for larger magnet thicknesses, thereby reducing the average longitudinal strain $\varepsilon_{zz}$ in the piezoelectric layer for a magnet thickness of 200 nm, as shown in Fig. 4 (a). In regime III, a larger axial strain is transferred to the piezoelectric layer when thicker magnets are used, due to an enhanced shear strain in thicker magnets for larger pillar diameters (App. C, Fig. ( 13)). Due to the Poisson effect, this results in an increase in the longitudinal strain $\varepsilon_{zz}$ for a magnet thickness of 200 nm (Fig. 4 (a)).

The surroundings of the magnetoelectric device play a crucial role in differential voltage generation, for example through surface or volume clamping effects. Since the ME composite is sandwiched between two electrodes, the strain transfer and the corresponding voltage generation also depend on the mechanical properties of the electrodes. In particular, we investigate the impact of the electrode metal on the strain transfer and the resulting differential voltage. Since the ME composite is sandwiched between two electrodes, the strain transfer and the corresponding voltage generation also depend on the mechanical properties of the electrodes. In Fig. 5 (a) the output voltage is plotted as function of the piezoelectric layer thickness for a pillar diameter of

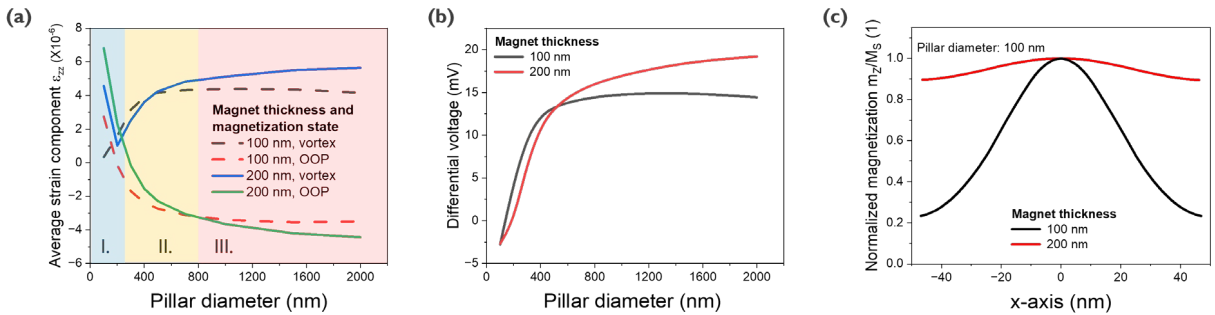

*Figure 4 (a) Average strain component $\varepsilon_{zz}$ in the piezoelectric layer for the ground (vortex) and uniform OOP magnetization states, for magnet thicknesses of 100 nm and 200 nm, respectively. (b)Differential voltage as a function of the pillar diameter for the two thickness of the Ni layer. (c) Normalized magnetization component $M_z/M_S$ for the two Ni thicknesses (100 nm and 200 nm) and a pillar of 100 nm diameter. The thicknesses of the piezoelectric layer and electrodes are set to 70 nm and 50 nm, respectively.*

500 nm and different electrode materials. As discussed above, bending occurs in the piezoelectric layer for pillars of these dimensions, as shown in Fig. 5(b) and (e). For thicker piezoelectric layers, a slight increase in bending occurs, leading to a small reduction in the differential voltage. Using stiffer electrodes, such as Ru, enhances strain transfer and thereby increases the generated differential voltage. Moreover, stiffer electrodes allow the strain to penetrate deeper into the piezoelectric layer, as illustrated in Fig. 5(c) and (f). For these stiffer electrodes, the bending effect at larger piezoelectric thicknesses is reduced, which decreases the difference in differential voltage between piezoelectric layer thicknesses of 200 nm and 400 nm. Clamping mechanically the whole device (top and bottom) leads to a drastically increase in the strain within the piezoelectric layer. However, because the pillar surface is clamped, axial displacement is suppressed. As a result, shear strain transfer is strongly reduced, and tensile strain is transferred directly to the piezoelectric layer by the compressive magnet. This effect is illustrated in Fig. 5(d) and (g), where tensile strain is distributed throughout the entire piezoelectric layer and bending is effectively suppressed. In the simulations, this boundary condition is realized by implementing fixed layers at the top of the top electrode and at the bottom of the bottom electrode. In practice, such mechanical clamping can be achieved by adding rigid encapsulation layers, for example SiN or diamond layers.

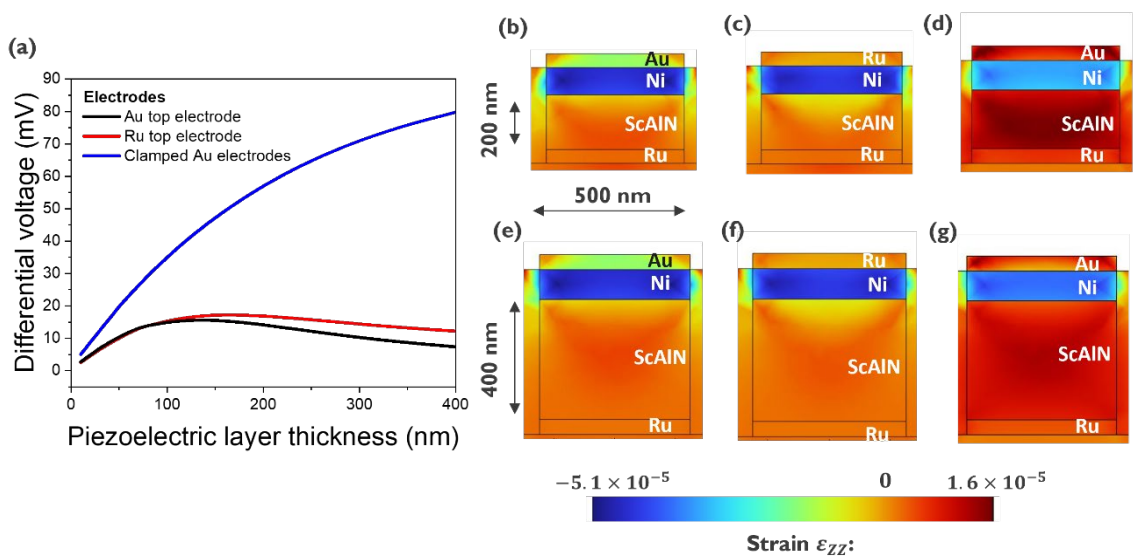

*Figure 5 (a) Differential voltage as a function of the piezoelectric layer thickness for different electrode materials (Au and Ru). Strain component $\varepsilon_{zz}$ for an OOP magnetization state in devices with a piezoelectric layer thickness of 200 nm (b-d) and 400 nm (e-g) with Au as the top electrode (b,e), Ru as the top electrode (c,f), and top–bottom clamped electrodes (d,g). The pillar diameter is set to 500 nm, and the magnet and electrode thicknesses are 100 nm and 50 nm, respectively.*

In the following, different magnetostrictive materials are investigated. As shown in Tab. 1, both FeGa and Tefenol-D exhibit stronger magnetoelastic coupling coefficients $B$ compared to Ni, therefore larger differential voltages are expected for devices based on these materials. Figure 6(a) display the differential voltage for devices based on FeGa layers of different thicknesses. Although the $B$-coefficient for FeGa is only approximately twice that of Ni, the resulting differential voltage increases by more than a factor of four. This can be explained by the large value Poisson ratio of FeGa. For an OOP magnetization state, the strain generated within the magnet is approximately twice that observed for Ni (see Fig. 5(b)). In addition, due to the higher Poisson ratio, a larger axial strain is induced, which enhances strain transfer through increased shear deformation. Fig. 6 (a) indicates that, depending on the device diameter, there exists an optimal piezoelectric layer thickness that yields the largest differential voltage. For devices with a diameter of 500 nm, the optimal piezoelectric layer thickness is approximately 100 nm, as shown in Fig. 6(b). Despite the fact that the largest average strain is transferred at small piezoelectric layer thicknesses (see Fig. 6(c) and (d)) the electric potential is maximized when the strain extends throughout the entire piezoelectric thickness, as shown in Fig. 6(e). For larger piezoelectric layer thicknesses, increased bending occurs, resulting in a strongly reduced average strain, as seen in Fig. 6(f). Whether increasing the magnet thickness leads to enhanced voltage generation depends on the piezoelectric layer thickness. At small piezoelectric thicknesses, a thicker magnet induces stronger bending in the piezoelectric layer due to increased competition between shear strain transfer and direct compression caused by magnet elongation. At larger piezoelectric thicknesses, increasing the magnet thickness can be beneficial, as bending is reduced because compression of the piezoelectric layer by the elongated magnet becomes the dominant mechanism. The inset in Fig. 6 (a) confirms that the magnet is in a vortex ground state for all studied thicknesses.

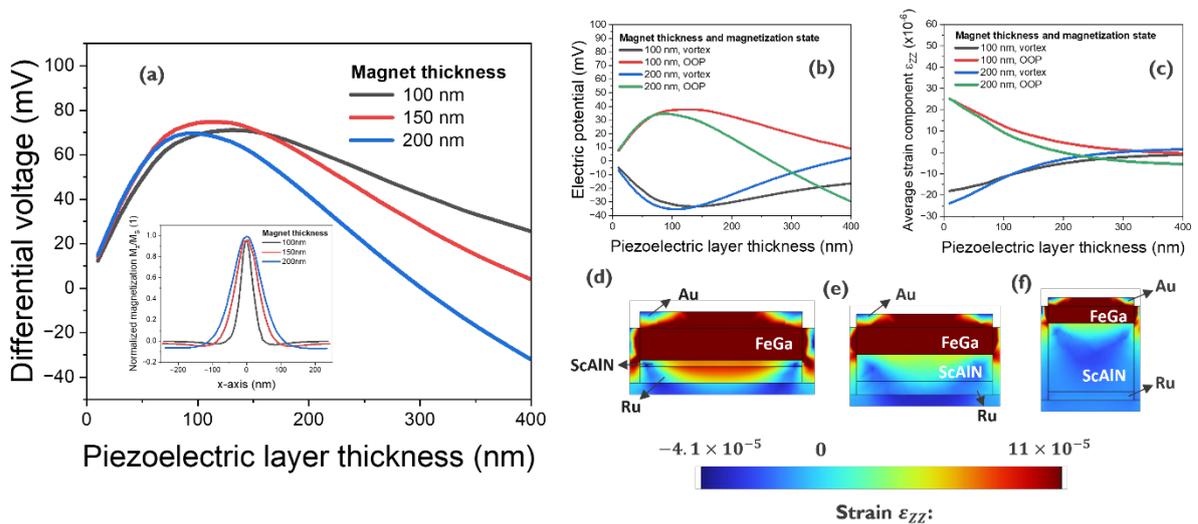

*Figure 6 Differential voltage and strain-transfer dependence on the magnet and piezoelectric layer thicknesses for a FeGa magnet and a device diameter of 500 nm. (a) Differential voltage as a function of piezoelectric layer thickness for different magnet thicknesses. The inset shows the normalized magnetization $M_z/M_S$ for three magnet thicknesses (100 nm, 150 nm and 200 nm). Average strain component in the piezoelectric layer (b) and the electric potential (c) for different magnet thicknesses and magnetization states. Panels (d-f) show the strain component $\varepsilon_{zz}$ for an OOP magnetization state in devices with a FeGa magnet of 100nm thickness and piezoelectric layer thicknesses of 20 nm (d), 100 nm (e), and 400 nm (f), respectively. The electrodes thicknesses are 50 nm.*

Terfenol-D is among the magnetostrictive materials with the highest strain output and exhibits an exceptionally large magnetoelastic coupling constant. Due to this strong magnetoelastic response, Terfenol-D can generate significantly larger strain under magnetic excitation compared to conventional ferromagnetic materials such as Ni or FeGa. To note that Terfenol-D is characterized by a cubic magnetocrystalline anisotropy, which should be carefully accounted for when modelling its magnetic ground state in nanoscale devices. Since cubic magnetocrystalline anisotropies cannot be directly implemented within the COMSOL simulation framework, the anisotropy of Terfenol-D is modelled using assumptions on the magnetic ground state as a function of the device diameter. For diameters up to 500 nm, a vortex magnetization state is assumed, while for diameters of 750 nm and above, the ground state is taken to be in-plane. This approach is validated by micromagnetic simulations carried out in MuMax3 including cubic anisotropy, as shown in Fig. 7(a) and (b). Since determining the exact transition between the two magnetic ground states is beyond the scope of this paper, devices with diameters between 500 nm and 750 nm were excluded from the analysis. Figure 7(b) displays the differential output voltage as a function of the pillar diameter for devices based on Ni, FeGa and Terfenol-D. As expected, Terfenol-D yields the highest differential voltage due to its exceptionally large magnetoelastic coupling constant $B$. Compared to Ni, $B$ is more than an order of magnitude larger, which is directly reflected in a differential voltage that is also more than ten times higher. As discussed above, in addition to the magnetoelastic coupling coefficients, the elastic properties of the magnetic material, such as the Poisson ratio, play a critical role in determining the voltage output. Note that $B$ is directly related to the magnetostriction coefficient $\lambda$ and the stiffness components of the magnet [61]. Consequently, materials exhibiting large values for $\lambda$ are not inherently optimal for voltage generation if their elastic stiffness coefficients (or, for isotropic materials, Young's modulus and Poisson ratio) do not favor efficient strain transfer to the piezoelectric layer. As the pillar diameter decreases, the differential voltage approaches zero and subsequently increases strongly with negative sign. This behavior is observed for all investigated materials, but is most pronounced for Terfenol-D. Similar to the case of Ni, the dominant strain-transfer mechanism for an OOP magnetization at small pillar diameters is direct compression resulting from magnet elongation (see Fig. 7(d)). As a result of this dominating strong direct compression at smaller pillar diameters, the differential voltage exceeds 100 mV for Terfenol-D for a pillar diameter of 100 mV. Since the efficiency of this direct strain transfer increases with decreasing pillar diameter, Fig. 7 (a) suggests that the output voltage increases further for sub 100-nm pillar diameters, illustrating a great potential for microelectronic applications. On the other hand, for increasing pillar diameter, strain transfer mediated by shear deformation becomes increasingly important and start counteracting the direct compression, as can be seen in Fig. 7(e). Beyond a certain diameter (~ 250 nm for Terfenol-D) shear strain transfer dominates, leading to a strong reduction of bending in the piezoelectric layer and the generation of a positive (tensile) strain, as illustrated in Fig. 7(f). Again, the obtained differential voltage saturates at increasing pillar diameters due to increasing surface clamping combined with an increasing shear strain transfer efficiency.

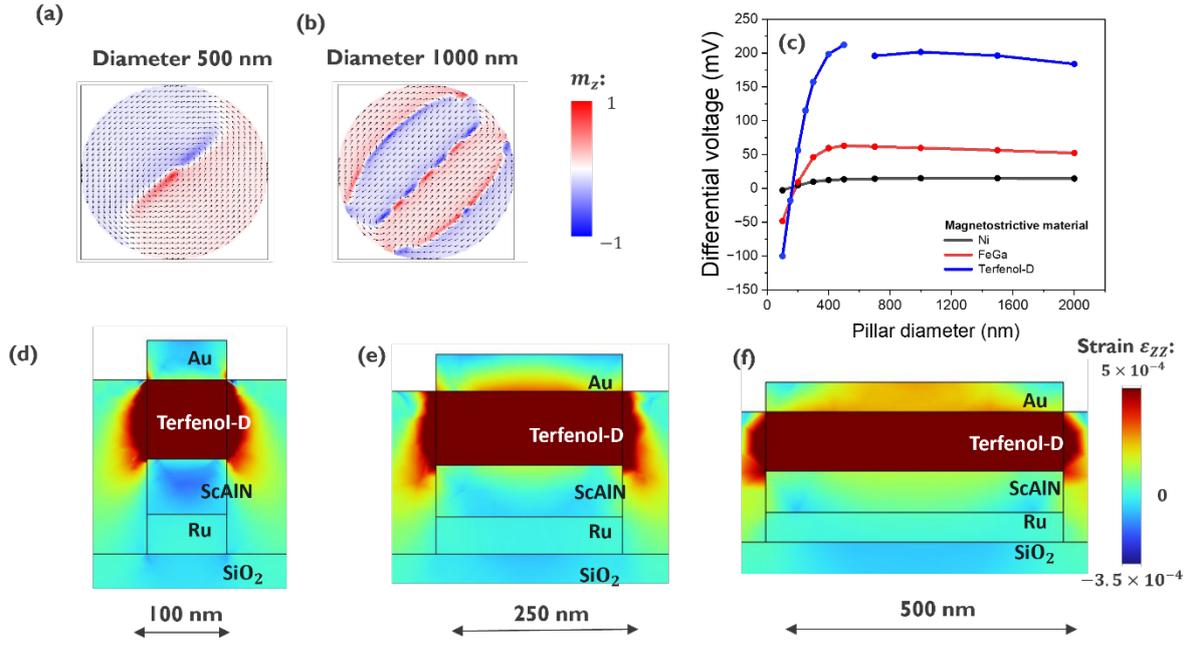

*Figure 7 Magnetic ground state of a Terfenol-D layer including cubic anisotropy for pillar diameters of 500 nm (a) and 1000 nm (b), respectively. (c) Differential voltage as a function of the pillar diameter for different magnetostrictive materials. Panels (d-f) show the strain component $\varepsilon_{zz}$ for an OOP magnetization state in devices with Terfenol-D as the magnetostrictive layer, for pillar diameters of 100 nm (d), 250 nm (e), and 500 nm (f). The magnet, piezoelectric layer, and electrode thicknesses are set to 100 nm, 70 nm and 50 nm, respectively.*

**IV. CONCLUSION**

In this work, we investigated the direct magnetoelectric (ME) effect in nanoscale composite ME devices, focusing on the strain transfer and corresponding achievable differential output voltage when the magnet is switched from its magnetic ground state to a uniform OOP configuration. Using a calibrated FEM model, we systematically varied geometrical and material parameters to gain insight into their impact on device performance and to establish design guidelines for scaled ME devices capable of generating large output voltages. Our results demonstrate that strain transfer can be highly efficient upon downscaling the pillar dimensions due to reduced surface clamping. Depending on the aspect ratio of the magnet and piezoelectric layers and the magnetization state, two dominant strain-transfer mechanisms are identified. For in-plane (IP) and vortex magnetization states, shear-mediated strain transfer is the dominating mechanism and becomes increasingly effective with larger diameter-to-thickness aspect ratios. In contrast, for OOP magnetization states, the strain transfer is dominated by direct compression or elongation of the piezoelectric layer, which is increasingly efficient for decreasing diameter-to-thickness ratio. On the other hand, for larger aspect ratios, shear strain transfer becomes dominant. However, the shear strain transfer is indirect for OOP magnetization states (because only axial strain is transferred by shear strain), and is therefore less efficient than the shear strain transfer in case of IP magnetization states. Overall, for increasing pillar diameters, the output voltage saturates due to enhanced surface clamping effects and increased shear-strain transfer efficiency. When scaling down ME devices, it is therefore essential to select sufficiently small aspect ratios to avoid competition between the two strain transfer mechanisms, which would otherwise induce bending in the piezoelectric layer and reduce the effective strain. Moreover, to ensure a maximal differential voltage between the magnetic ground state and the OOP state, the OOP component in the ground-state magnetization should be minimized. This can be achieved through appropriate aspect ratio selection or by choosing magnetic materials with large exchange stiffness or pronounced IP anisotropy. In addition, the magnetoelastic coupling can be further enhanced by employing stiffer electrode materials or by introducing mechanical clamping layers, improving the strain transfer to the piezoelectric layer. Magnetostrictive materials such as FeGa and Terfenol-D are particularly promising due to their favorable magnetostrictive and elastic properties, enabling output voltages exceeding 100 mV for nanoscale pillars with diameters approaching 100 nm. Combining mechanical clamping with optimized magnetic and piezoelectric layer thicknesses can further strengthen the magnetoelastic coupling and substantially increase the differential voltage. Overall, this work provides detailed insight into the strain transfer mechanisms in nanoscale ME devices and demonstrates the feasibility of achieving large output voltages at sub-100 nm dimensions, thereby highlighting the potential for next-generation energy efficient memory, logic and magnetic sensing technologies.

## V.   ACKNOWLEDGMENT

E.V.M. gratefully acknowledges financial support from the Research foundation – Flanders (FWO) through grant No. 1SH4Q24N. This work has been supported by imec's industrial affiliate program on Exploratory Logic.

**Appendix A: Axial strain transfer for uniform IP and OOP magnetization**

The axial load transfer between different perfectly bonded layers is described by the shear lag model, which is a theoretical approach investigating the interfacial stress transfer mechanism in fibrous and layered composites with an adhesive layer [69, 72, 73]. This model states that, when a layer is loaded axially in tension or compression, this load is transferred in the adhesive layer through shear stresses that are distributed along the length of the bond [70]. This shear lag model provides some insight in the strain transfer mechanisms in the ME device when the magnet is axially compressed or elongated by magnetostriction. Fig. 8 (a) shows the normal axial strain $\varepsilon_{xx}$ and the shear strain component $\varepsilon_{xz}$ along the magnet-piezoelectric layer interface when the magnetization is in an IP magnetization state. Due to the large axial magnetostrictive strain arising in the magnet by the IP magnetization, a mismatch in displacement with the piezoelectric layer is generated. This causes shear strain $\varepsilon_{xz}$ to arise at the interface, which is largest near the pillar edges, as seen in Fig. 8 (a). To counteract this shear strain, an axial strain $\varepsilon_{xx}$ is generated in the piezoelectric layer at the interface, which decays towards the pillar center. Similar strain profiles arise at the interface in case of an OOP magnetization distribution. However, the strain is smaller in magnitude, since the axial strain $\varepsilon_{xx}$ in the magnet is arising mainly from the magnetostrictive longitudinal strain $\varepsilon_{zz}$ through the Poison Effect. Hence, the longitudinal strain $\varepsilon_{zz}$ is indirectly transferred by shear strain transfer through the Poisson effect.

Assuming that the magnet's thickness remains constant, the transferred strain penetrates the piezoelectric depth depending on the pillar's diameter. Fig. 8 (b) shows how the transferred axial strain $\varepsilon_{xx}$ in the center of the piezoelectric layer penetrates into the piezoelectric layer. For a pillar diameter of 100 nm, the transferred strain decays quickly to zero, whereas for a pillar diameter of 500 nm, $\varepsilon_{xx}$ penetrates deeper into the piezoelectric layer, only reaching zero towards the interface between the piezoelectric layer and the bottom electrode. For a pillar diameter of 2000 nm, $\varepsilon_{xx}$ penetrates through the entire piezoelectric layer with limited decay. Note that the magnitude of the strain at the magnet-piezoelectric interface decreases with increasing pillar diameter. This is because the average magnetostrictive strain in the magnet decreases with increasing pillar diameter, which can be attributed to increasing surface clamping effects.

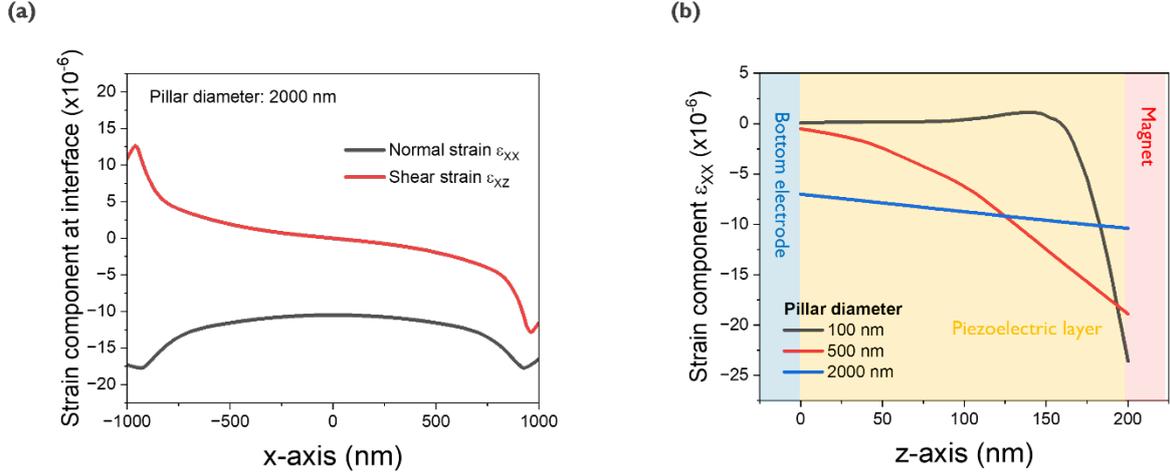

*Figure 8 The strain components in the ME device from Fig. 3, using a Ni magnet with a thickness of 100 nm, for an IP magnetization state. (a) Normal and shear strain component for an in-plane magnetization distribution at the interface of the magnetostrictive and piezoelectric layer for a pillar diameter of 2000 nm. (b) Strain component $\varepsilon_{xx}$ in the center of a 200 nm thick piezoelectric layer. The origin of the z-axis is defined at the interface between the bottom electrode and the piezoelectric layer.*

**Appendix B: Strain transfer efficiency**

The strain-transfer efficiency is defined as the ratio between the magnitude of the average strain present in the piezoelectric layer and the magnitude of the average strain generated in the magnet by magnetostriction. Figures 9 (a) and (c) show the average strain $\varepsilon_{zz}$ in the magnet and piezoelectric laye, as well as the resulting strain-transfer efficiency, for a uniform IP magnetization state, whereas Figs. 9 (b) and (d) present the corresponding results for a uniform OOP magnetization state. In both cases, the strain generated in the magnet decreases with increasing pillar diameter due to enhanced surface-clamping effects at larger pillar areas. Conversely, the average strain transferred to the piezoelectric layer increases with increasing pillar diameter. Together, these trends explain the increase in strain-transfer efficiency with pillar diameter. Furthermore, Figs. 9(c) and (d) show that the strain-transfer efficiency is significantly smaller for an OOP magnetization state. However, Figs. 9(a) and (b) indicate that the average strain in the piezoelectric layer is similar for both IP and OOP magnetization states. This difference in efficiency arises from the larger strain that is generated in the magnet for an OOP magnetization state compared to an IP state. Because the magnetostrictive material is isotropic, this difference is purely due to geometrical factors and surface-clamping effects. The difference in strain-transfer behavior between IP and OOP magnetization states can be understood as follows. For an IP magnetization state, magnetostriction directly generates axial strain in the magnet, which is transferred to the piezoelectric layer through shear strain at the interface (see App. A). In contrast, for an OOP magnetization state, the longitudinal magnetostrictive strain is first converted into axial strain via the Poisson effect before being transferred through shear strain. This indirect transfer mechanism becomes dominant at larger pillar diameters, resulting in a lower strain-transfer efficiency for OOP magnetization states compared to IP states. On the other hand, for small pillar diameters, direct strain transfer is the dominant mechanism for OOP magnetization states. Since this direct transfer becomes more efficient at smaller aspect ratios, the strain-transfer efficiency for OOP magnetization states increases as the pillar diameter decreases.

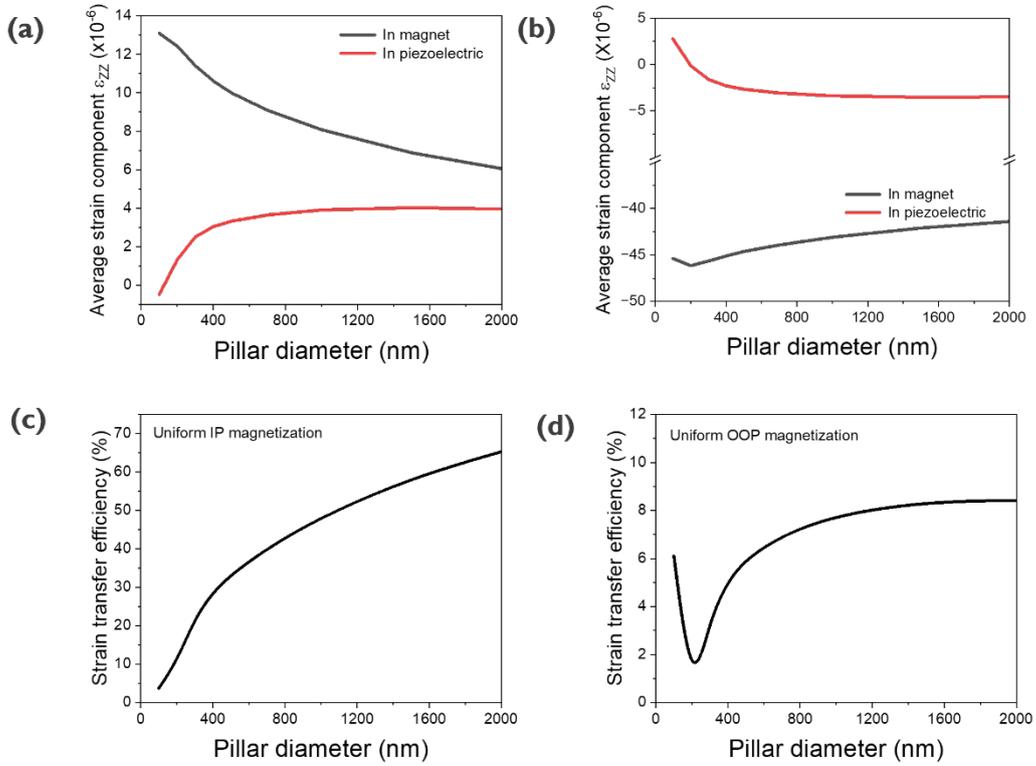

*Figure 9 Impact of pillar diameter on the average strain developed in the magnet and the piezoelectric layer for (a) a uniform IP magnetization state and (b) a uniform OOP magnetization state. Dependence of the pillar diameter on the strain-transfer efficiency for (c) a uniform IP magnetization state and (d) a uniform OOP magnetization state. In all cases, the ME device shown in Fig. 3 is used, with a Ni magnet of 100 nm thickness.*

**Appendix C: Normal and shear strain distributions**

The distribution of the shear strain component $\varepsilon_{xz}$ is illustrated in Fig. 10 for different pillar diameters, for both a vortex and a uniform OOP magnetization state. The non-uniform magnetization of the vortex state produces a non-uniform strain distribution throughout the magnetic volume, leading to shear strain across the entire magnet–piezoelectric interface, as illustrated in Fig. 10(a). In contrast, for the uniform OOP magnetization state, the axial strain in the magnet is much more uniform, and shear strain appears predominantly at the interface edges (consistent with the shear-lag model described in App. A). Figure 8(b) shows that, for a uniform IP magnetization state, the strain penetration depth into the piezoelectric layer increases with pillar diameter. A similar trend is found for the vortex and uniform OOP magnetization states: Fig. 11 demonstrates that an increase in pillar diameter enhances shear-strain transfer into the piezoelectric layer, leading to a larger normal strain deeper inside the piezoelectric film. Consequently, increasing pillar diameter results in a higher strain-transfer efficiency, as previously discussed in Figs. 9(c) and (d).

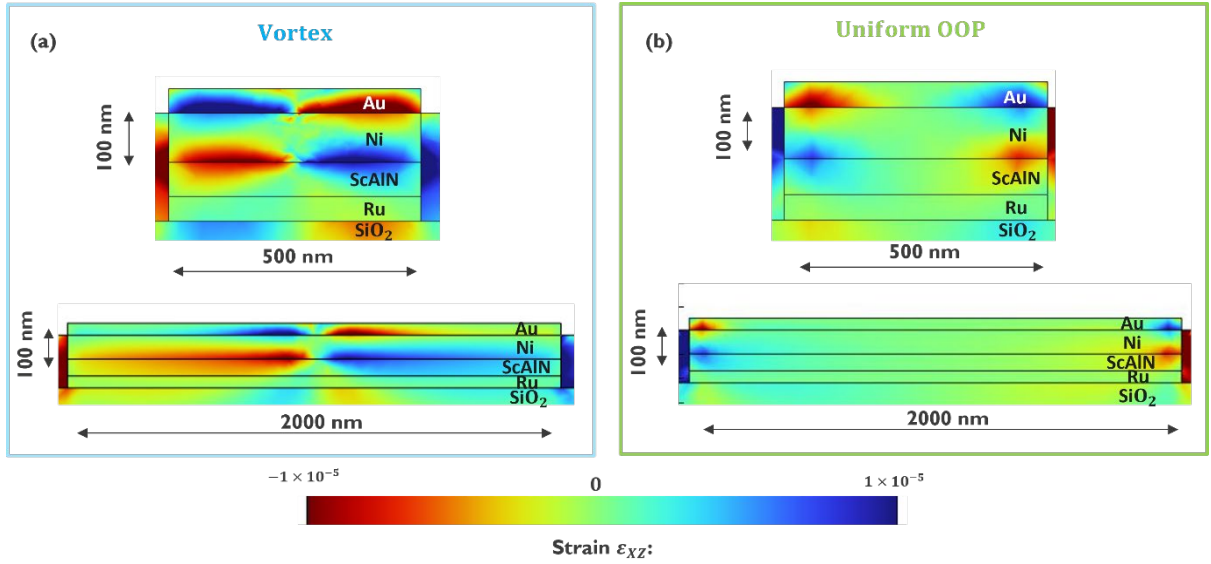

*Figure 10 Shear strain component $\varepsilon_{xz}$ for the pillar of Fig. 4, shown for magnet thicknesses of 100 nm and pillar diameters of 500 nm and 2000 nm, respectively. The piezoelectric layer and electrode thicknesses are set to 70 nm and 50 nm, respectively. The shear strain distributions are illustrated for (a) a vortex magnetization state and (b) a uniform OOP magnetization state.*

Figure 12 presents the longitudinal strain distribution $\varepsilon_{zz}$ for different pillar diameters for an OOP magnetization state and a magnet thickness of 200 nm. For small pillar diameters (Fig. 12(a)), tensile longitudinal strain develops in the piezoelectric layer because the magnet compresses along the z-axis and directly elongates the piezoelectric material. For large pillar diameters (Fig. 12(c)), compressive longitudinal strain is observed due to axial load transfer through shear-mediated strain transfer. At intermediate pillar diameters (Fig. 12(b)), both tensile and compressive regions coexist.

Figure 12 also illustrates the enhanced strain transfer resulting from the larger magnet thickness (compare with Figs. 3(g), (h), and (i)). This enhancement arises from both stronger direct strain transfer and increased shear-strain transfer. To demonstrate that the increase in normal strain at larger pillar diameters is due to increased shear-strain transfer, Fig. 13 plots the average magnitude of the shear strain within the piezoelectric layer as a function of pillar diameter. Indeed, Fig. 13 shows a clear increase in shear-strain transfer with increasing pillar diameter. For pillar diameters exceeding approximately 1000 nm, the average shear strain decreases again because shear occurs only at the pillar perimeter, which represents a smaller fraction of the total volume as device diameter increases. Moreover, Fig. 13 shows that, for pillar diameters around 400 nm, larger magnet thicknesses generate larger shear strain in the piezoelectric layer. This explains why Fig. 4(a) shows increased longitudinal strain for a magnet thickness of 200 nm at larger pillar diameters. However, in Fig. 4(a), a larger average normal strain for the 200 nm magnet appears only at diameters above ~700 nm for the OOP magnetization state, despite the increase in shear strain already occurring above 400 nm. This discrepancy arises because increased bending is present at intermediate diameters (400-700 nm) for the thicker magnet, which reduces the average normal strain in the piezoelectric layer.

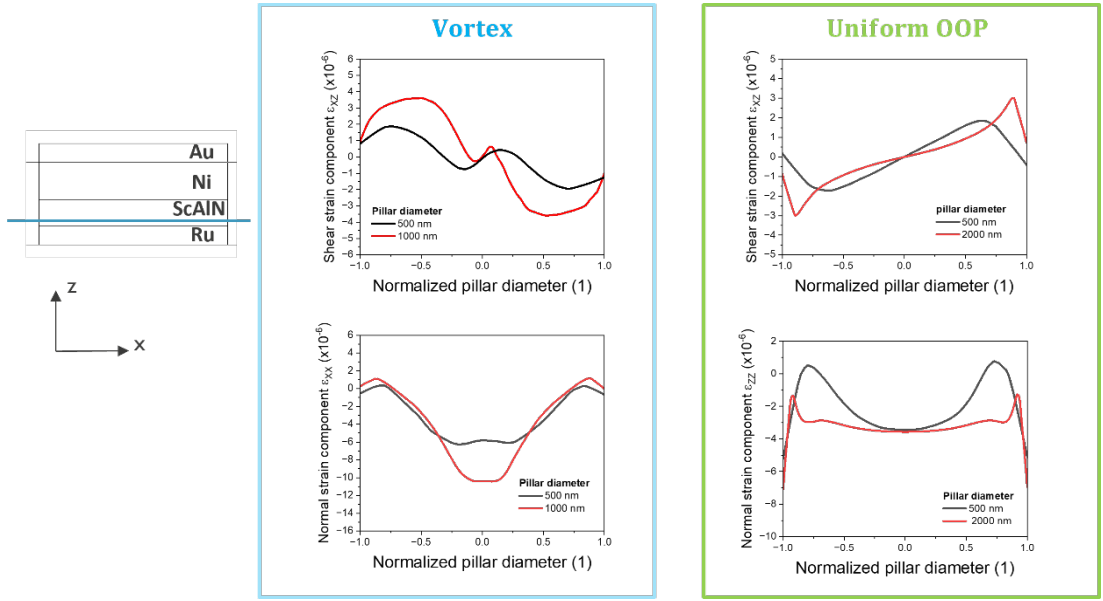

*Figure 11 Shear strain ε$_{xz}$ and normal strain component ε$_{xx}$ for different pillar diameters for vortex and OOP magnetization states. The data is taken on the cut line at the lower edge of the piezoelectric layer, as seen in the figure.*

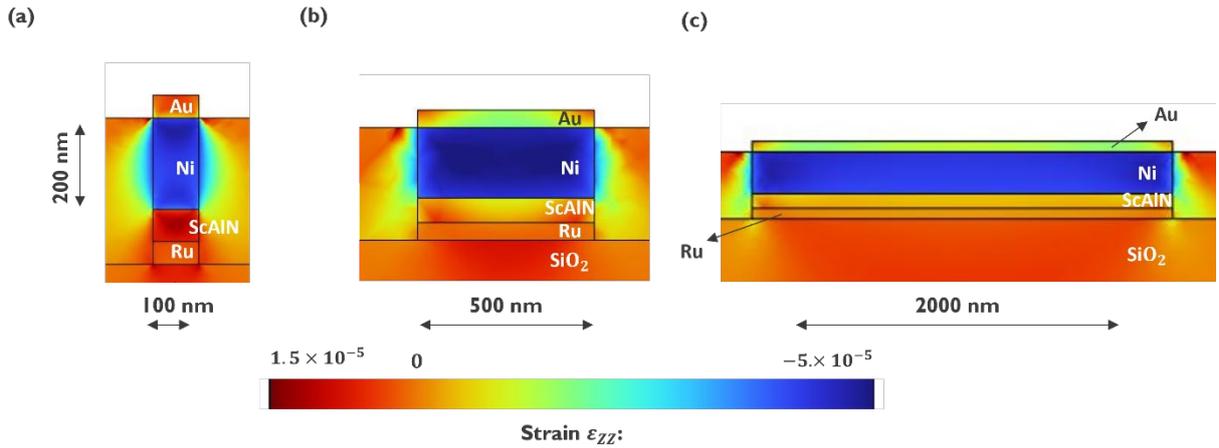

*Figure 12 Strain component ε$_{zz}$ for an OOP magnetization state for the pillar from Fig. 4, shown for a magnet thickness of 200 nm and pillar diameters of 100 nm (a), 500 nm (b), and 2000 nm (c). The piezoelectric layer and electrode thickness are set to 70 nm and 50 nm, respectively.*

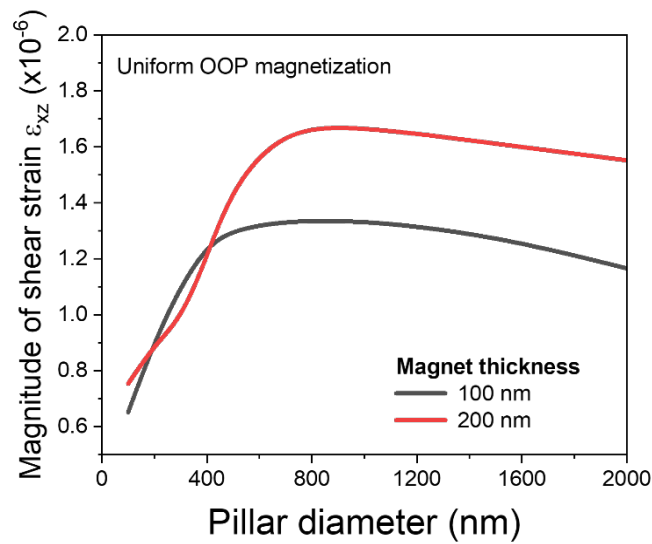

*Figure 13 Average magnitude of the shear strain ε$_{xz}$ in the piezoelectric layer as a function of the pillar diameter for magnet thicknesses of 100 nm and 200 nm, and an OOP magnetization state.*

# References

[1] Jia-Mian Hu, Long-Qing Chen, and Ce-Wen Nan. Multiferroic heterostructures integrating ferroelectric and magnetic materials. *Advanced materials (Weinheim)*, 28(1):15–39, 2016.

[2] N. A. Spaldin and R. Ramesh. Advances in magnetoelectric multiferroics. *Nature materials*, 18(3):203–212, 2019.

[3] Xianfeng Liang, Huaihao Chen, and Nian X. Sun. Magnetoelectric materials and devices. *APL materials*, 9(4):041114–041114–27, 2021.

[4] Xianfeng Liang, Alexei Matyushov, Patrick Hayes, Viktor Schell, Cunzheng Dong, Huaihao Chen, Yifan He, Alexandria Will-Cole, Eckhard Quandt, Pedro Martins, Jeffrey McCord, Marisa Medarde, Senentxu Lanceros-Mendez, Sebastiaan van Dijken, Nian X. Sun, and Jordi Sort. Roadmap on magnetoelectric materials and devices. *IEEE Transactions on Magnetics*, 57(8):1–57, 2021.

[5] Albert Fert, Ramamoorthy Ramesh, Vincent Garcia, Fèlix Casanova, and Manuel Bibes. Electrical control of magnetism by electric field and current-induced torques. *Reviews of Modern Physics*, 96(1):1–, 2024.

[6] Ayan Kumar Biswas, Hasnain Ahmad, Jayasimha Atulasimha, and Supriyo Bandyopadhyay. Experimental demonstration of complete $180°$ reversal of magnetization in isolated co nano-magnets on a pmn–pt substrate with voltage generated strain. *Nano letters*, 17(6):3478–3484, 2017.

[7] Xu Li, Dorinamaria Carka, Cheng-yen Liang, Abdon E. Sepulveda, Scott M. Keller, Pe-dram Khalili Amiri, Gregory P. Carman, and Christopher S. Lynch. Strain-mediated $180°$ perpendicular magnetization switching of a single domain multiferroic structure. *Journal of applied physics*, 118(1):14101–, 2015.

[8] Gajanan Pradhan, Federica Celegato, Alessandro Magni, Marco Coisson, Gabriele Barrera, Paola Rizzi, and Paola Tiberto. Electric field control of magnetization reversal in fega/pmn-pt thin films. *JPhys materials*, 7(1):15016–, 2024.

[9] Peter B. Meisenheimer, Steve Novakov, Nguyen M. Vu, and John T. Heron. Perspective: Magnetoelectric switching in thin film multiferroic heterostructures. *Journal of applied physics*, 123(24), 2018.

[10] Ramamoorthy Ramesh, Sayeef Salahuddin, Suman Datta, Carlos H. Diaz, Dmitri E. Nikonov, Ian A. Young, Donhee Ham, Meng-Fan Chang, Win-San Khwa, Ashwin Sanjay Lele, Christian Binek, Yen-Lin Huang, Yuan-Chen Sun, Ying-Hao Chu, Bhagwati Prasad, Michael Hoffmann, Jia-Mian

Hu, Zhi (Jackie) Yao, Laurent Bellaiche, Peng Wu, Jun Cai, Joerg Appenzeller, Supriyo Datta, Kerem Y. Camsari, Jaesuk Kwon, Jean Anne C. Incorvia, Inge Asselberghs, Florin Ciubotaru, Sebastien Couet, Christoph Adelmann, Yi Zheng, Aaron M. Lindenberg, Paul G. Evans, Peter Ercius, and Iuliana P. Radu. Roadmap on low-power electronics. *APL materials*, 12(9):099201–099201–73, 2024.

[11] Sasikanth Manipatruni, Dmitri E. Nikonov, Chia-Ching Lin, Tanay A. Gosavi, Huichu Liu, Bhagwati Prasad, Yen-Lin Huang, Everton Bonturim, Ramamoorthy Ramesh, and Ian A. Young. Scalable energy-efficient magnetoelectric spin–orbit logic. *Nature (London)*, 565(7737):35–42, 2019.

[12] R. Ramesh and Nicola A. Spaldin. Multiferroics: progress and prospects in thin films. *Nature materials*, 6(1):21–29, 2007.

[13] Gustau Catalan and James F. Scott. Physics and applications of bismuth ferrite. *Advanced materials (Weinheim)*, 21(24):2463–2485, 2009.

[14] J. T. Heron, D. G. Schlom, and R. Ramesh. Electric field control of magnetism using BiFeO3-based heterostructures. *Applied Physics Reviews*, 1(2):021303, 04 2014.

[15] Julia A. Mundy, Charles M. Brooks, Megan E. Holtz, Jarrett A. Moyer, Hena Das, Ale-jandro F. Rébola, John T. Heron, James D. Clarkson, Steven M. Disseler, Zhiqi Liu, Alan Farhan, Rainer Held, Robert Hovden, Elliot Padgett, Qingyun Mao, Hanjong Paik, Rajiv Misra, Lena F. Kourkoutis, Elke Arenholz, Andreas Scholl, Julie A. Borchers, William D. Ratcliff, Ramamoorthy Ramesh, Craig J. Fennie, Peter Schiffer, David A. Muller, and Dar-rell G. Schlom. Atomically engineered ferroic layers yield a room-temperature magnetoelectric multiferroic. *Nature (London)*, 537(7621):523–527, 2016.

[16] Nicola A Hill. Why are there so few magnetic ferroelectrics? *The journal of physical chemistry. B*, 104(29):6694–6709, 2000.

[17] Claude Ederer and Nicola A Spaldin. Weak ferromagnetism and magnetoelectric coupling in bismuth ferrite. *Phys. Rev. B*, 71:060401(R), 2004.

[18] Daniele Narducci, Xiangyu Wu, Isabella Boventer, Jo De Boeck, Abdelmadjid Anane, Paolo Bortolotti, Christoph Adelmann, and Florin Ciubotaru. Magnetoelectric coupling in ba:pb(zr,ti)o3/co40fe40b20 nanoscale waveguides studied by propagating spin-wave spec-troscopy. *Applied Physics Letters*, 124(18), 2024.

[19] Yaojin Wang, Jiefang Li, and D. Viehland. Magnetoelectrics for magnetic sensor applications: status, challenges and perspectives. *Materials today (Kidlington, England)*, 17(6):269–275, 2014.

[20] A. A. Amirov, D. M. Yusupov, A. M. Ismailov, and N. Z. Abdulkadirova. Direct magne-toelectric effect in ni–pzt–pt layered multiferroic composites. *Surface investigation, x-ray, synchrotron and neutron techniques*, 12(2):336–338, 2018.

[21] William Paul Flynn, Sean Garnsey, Amar S. Bhalla, and Ruyan Guo. Finite element analysis of strain-mediated direct magnetoelectric coupling in multiferroic nanocomposites for material jetting fabrication of tunable devices. *Journal of composites science*, 9(5):228–, 2025.

[22] P Martins, X Moya, L C Phillips, S Kar-Narayan, N D Mathur, and S Lanceros-Mendez. Linear anhysteretic direct magnetoelectric effect in ni0.5zn0.5fe2o4/poly(vinylidene fluoride-trifluoroethylene) 0-3 nanocomposites. *Journal of Physics. D, Applied Physics*, 44(48):482001–, 2011.

[23] Junyi Zhai, Zengping Xing, Shuxiang Dong, Jiefang Li, and Dwight Viehland. Magnetoelectric laminate composites: An overview. *Journal of the American Ceramic Society*, 91(2):351–358, 2008.

[24] Federica Luciano, Erika Giorgione, Emma Van Meirvenne, Andrei Galan, Ilaria Marzorati, Arne De Coster, Dominika Wysocka, Bart Sorée, Stefan De Gendt, Florin Ciubotaru, and Christoph Adelmann. Magnetic-field induced charge accumulation in scalable magnetoelectric pvdf-trfe/ni composite devices. *Advanced Physics Research*, 2025.

[25] R. Ghosh, A. Barik, M. R. Sahoo, Sweta Tiwary, P. D. Babu, S. D. Kaushik, and P. N. Vishwakarma. Display of converse and direct magnetoelectric effect in double perovskite layfe2o6. *Journal of Applied Physics*, 132(22), 2022.

[26] Morgan Trassin. Low energy consumption spintronics using multiferroic heterostructures. *Journal of physics. Condensed matter*, 28(3):33001–, 2016.

[27] Sasikanth Manipatruni, Dmitri E. Nikonov, and Ian A. Young. Beyond cmos computing with spin and polarization. *Nature physics*, 14(4):338–343, 2018.

[28] Abdulqader Mahmoud, Florin Ciubotaru, Frederic Vanderveken, Andrii V. Chumak, Said Hamdioui, Christoph Adelmann, and Sorin Cotofana. Introduction to spin wave computing. *Journal of applied physics*, 128(16), 2020.

[29] John T. Heron and Tony Chiang. Magnetoelectrics and multiferroics: Materials and opportu-nities for energy-efficient spin-based memory and logic. *MRS bulletin*, 46(10):938–945, 2021.

[30] Shanfei Zhang, Zhuofan Li, Yizhuo Xu, and Bin Su. Flexible magnetoelectric systems: Types, principles, materials, preparation and application. *Applied Physics Reviews*, 11(4):041321, 11 2024.

[31] Shashi Poddar, Pedro de Sa, Ronggang Cai, Laurent Delannay, Bernard Nysten, Luc Piraux, and Alain M Jonas. Room-temperature magnetic switching of the electric polarization in ferroelectric

nanopillars. *ACS nano*, 12(1):576–584, 2018.

[32] J.A.C. Incorvia, T.P. Xiao, Zogbi N., A. Naeemi, C. Adelmann, F. Catthoor, M. Tahoori, F. Casanova, M. Becherer, G. Prenat, and S. Couet. Spintronics for achieving system-level energy-efficient logic. *Nature Reviews Electrical Engineering*, 1:700–713, 2024.

[33] Bhagwati Prasad, Yen-Lin Huang, Rajesh V. Chopdekar, Zuhuang Chen, James Steffes, Sujit Das, Qian Li, Mengmeng Yang, Chia-Ching Lin, Tanay Gosavi, Dmitri E. Nikonov, Zi Qiang Qiu, Lane W. Martin, Bryan D Huey, Ian Young, Jorge Íñiguez, Sasikanth Manipatruni, and Ramamoorthy Ramesh. Ultralow voltage manipulation of ferromagnetism. *Advanced materials (Weinheim)*, 32(28):e2001943–n/a, 2020.

[34] K. Lefki and G. J. M. Dormans. Measurement of piezoelectric coefficients of ferroelectric thin films. *Journal of applied physics*, 76(3):1764–1767, 1994.

[35] I. Fina, N. Dix, J. M. Rebled, P. Gemeiner, X. Martí, F. Peiró, B. Dkhil, F. Sánchez, L. Fàbrega, and J. Fontcuberta. The direct magnetoelectric effect in ferroelectric-ferromagnetic epitaxial heterostructures. *Nanoscale*, 5(17):8037–8044, 2013.

[36] Sen Zhang, Yonggang Zhao, Xia Xiao, Yizheng Wu, Syed Rizwan, Lifeng Yang, Peisen Li, Jiawei Wang, Meihong Zhu, Huiyun Zhang, Xiaofeng Jin, and Xiufeng Han. Giant electrical modulation of magnetization in co40fe40b20/pb(mg1/3nb2/3)0.7ti0.3o3(011) heterostructure. *Scientific reports*, 4(1):3727–, 2014.

[37] Haribabu Palneedi, Deepam Maurya, Gi-Yeop Kim, Shashank Priya, Suk-Joong L. Kang, Kwang-Ho Kim, Si-Young Choi, and Jungho Ryu. Enhanced off-resonance magnetoelectric response in laser annealed pzt thick film grown on magnetostrictive amorphous metal sub-strate. *Applied Physics Letters*, 107(1), 2015.

[38] Haribabu Palneedi, Hong Goo Yeo, Geon-Tae Hwang, Venkateswarlu Annapureddy, Jong-Woo Kim, Jong-Jin Choi, Susan Trolier-McKinstry, and Jungho Ryu. A flexible, high-performance magnetoelectric heterostructure of (001) oriented pb(zr 0.52 ti 0.48 )o 3 film grown on ni foil. *APL Materials*, 5(9):96111–, 2017.

[39] J. Irwin, S. Lindemann, W. Maeng, J. J. Wang, V. Vaithyanathan, J. M. Hu, L. Q. Chen, D. G. Schlom, C. B. Eom, and M. S. Rzchowski. Magnetoelectric coupling by piezoelectric tensor design. *Scientific reports*, 9(1):19158–9, 2019.

[40] Hongzhi Wu, Ruiying Luo, Zhuofan Li, Yujia Tian, Jiayi Yuan, Bin Su, Kun Zhou, Chunze Yan, and Yusheng Shi. Additively manufactured flexible liquid metal–coated self-powered magnetoelectric sensors with high design freedom. *Advanced materials (Weinheim)*, 36(34):e2307546–

n/a, 2024.

[41] Jun Kyu Choe, Suntae Kim, Ah-young Lee, Cholong Choi, Jae-Hyeon Cho, Wook Jo, My-oung Hoon Song, Chaenyung Cha, and Jiyun Kim. Flexible, biodegradable, and wireless magnetoelectric paper for simple in situ personalization of bioelectric implants (adv. mater. 18/2024). *Advanced materials (Weinheim)*, 36(18), 2024.

[42] Guohua Dong, Tian Wang, Haixia Liu, Yijun Zhang, Yanan Zhao, Zhongqiang Hu, Wei Ren, Zuo-Guang Ye, Keqing Shi, Ziyao Zhou, Ming Liu, and Jingye Pan. Strain-induced magnetoelectric coupling in fe3o4/batio3 nanopillar composites. *ACS applied materials and interfaces*, 14(11):13925–13931, 2022.

[43] Marijn W. van de Putte, Dmytro Polishchuk, Nicolas Gauquelin, Johan Verbeeck, Gertjan Koster, and Mark Huijben. Control of magnetic shape anisotropy by nanopillar dimensionality in vertically aligned nanocomposites. *ACS applied electronic materials*, 6(5):3695–3703, 2024.

[44] D Tierno, F Ciubotaru, R Duflou, M Heyns, IP Radu, and C Adelmann. Strain coupling optimization in magnetoelectric transducers. *Microelectronic Engineering*, 187-188:144–147, 2018.

[45] Federica Luciano, Emma Van Meirvenne, Ephraim Spindler, Philipp Pirro, Bart Sorée, Math-ias Weiler, Stefan De Gendt, Florin Ciubotaru, and Christoph Adelmann. Charge accumula-tion by direct magnetoelectric effect in scaln/ni nanoscale devices, 2025.

[46] Diogo C. Vaz, Chia-Ching Lin, John J. Plombon, Won Young Choi, Inge Groen, Isabel C. Arango, Andrey Chuvilin, Luis E. Hueso, Dmitri E. Nikonov, Hai Li, Punyashloka Debashis, Scott B. Clendenning, Tanay A. Gosavi, Yen-Lin Huang, Bhagwati Prasad, Ramamoorthy Ramesh, Aymeric Vecchiola, Manuel Bibes, Karim Bouzehouane, Stephane Fusil, Vincent Garcia, Ian A. Young, and Fèlix Casanova. Voltage-based magnetization switching and reading in magnetoelectric spin-orbit nanodevices. *Nature communications*, 15(1):1902–9, 2024.

[47] Takuma Sato, Weichao Yu, Simon Streib, and Gerrit E. W Bauer. Dynamic magnetoelastic boundary conditions and the pumping of phonons. *Phys. Rev. B*, 104:014403, 2021.

[48] Frederic Vanderveken, Jeroen Mulkers, Jonathan Leliaert, Bartel Van Waeyenberge, Bart Soree, Odysseas Zografos, Florin Ciubotaru, and Christoph Adelmann. Confined magnetoe-lastic waves in thin waveguides. *Phys. Rev. B*, 103:054439, 2021.

[49] Kei Yamamoto, Weichao Yu, Tao Yu, Jorge Puebla, Mingran Xu, Sadamichi Maekawa, and Gerrit Bauer. Non-reciprocal pumping of surface acoustic waves by spin wave resonance. *Journal of the Physical Society of Japan*, 89(11):113702–, 2020.

[50] Weichao Yu and Jiang Xiao. Spin dynamics simulation based on micromagnetic models. https://cn.comsol.com/paper/3c2a96daab4a69c6bbe7a6a3f91e7968-67443, 2018. Accessed: 2025–10-23.

[51] Zhengping Sun, Lei Li, Guolai Yang, and Liqun Wang. Micromagnetic simulation of nd-fe-b demagnetization behavior in complex environments. *Journal of magnetism and magnetic materials*, 589:171555–, 2024.

[52] Arne Vansteenkiste, Jonathan Leliaert, Mykola Dvornik, Mathias Helsen, Felipe Garcia-Sanchez, and Bartel Van Waeyenberge. The design and verification of Mumax3. *AIP Advances*, 4(10):107133, 2014.

[53] Göran Engdahl. Handbook of giant magnetostrictive materials. Elsevier Science and Technology, United States, 1999.

[54] Daining. Fang, issuing body. Tsinghua University Press, and Jinxi. Liu. *Fracture Mechanics of Piezoelectric and Ferroelectric Solids*. Springer Berlin Heidelberg, Berlin, Heidelberg, 1st ed. 2013. edition, 2013.

[55] Miguel A Caro, Siyuan Zhang, Tommi Riekkinen, Markku Ylilammi, Michelle A Moram, Olga Lopez-Acevedo, Jyrki Molarius, and Tomi Laurila. Piezoelectric coefficients and spontaneous polarization of scaln. *Journal of physics. Condensed matter*, 27(24):245901–245901, 2015.

[56] Dirk Sander, Stuart S.P. Parkin, and J. M. D. Coey. Magnetostriction and magnetoelastic-ity. In *Handbook of Magnetism and Magnetic Materials*, volume 1, pages 549–593. Springer International Publishing, Cham, 2021.

[57] M. D. Kuz'min, K. P. Skokov, L. V. B. Diop, I. A. Radulov, and O. Gutfleisch. Exchange stiffness of ferromagnets. *European physical journal plus*, 135(3):301–, 2020.

[58] Bong-Hi Shin and Young-Woo Park. Proof-of-concept of magnetic wheel-based magnetostric-tive energy harvester. *Journal of the Korean Society for Precision Engineering*, 32:483–490, 05 2015.

[59] S. Datta, J. Atulasimha, C. Mudivarthi, and A.B. Flatau. Stress and magnetic field-dependent young's modulus in single crystal iron–gallium alloys. *Journal of magnetism and magnetic materials*, 322(15):2135–2144, 2010.

[60] Holly Schurter and Alison Flatau. Elastic properties and auxetic behavior of galfenol for a range of compositions. *Proc SPIE*, 6929, 03 2008.

[61] P. B. Meisenheimer, R. A. Steinhardt, S. H. Sung, L. D. Williams, S. Zhuang, M. E. Nowakowski, S. Novakov, M. M. Torunbalci, B. Prasad, C. J. Zollner, Z. Wang, N. M. Daw-ley, J. Schubert, A. H. Hunter, S. Manipatruni, D. E. Nikonov, I. A. Young, L. Q. Chen,

J. Bokor, S. A. Bhave, R. Ramesh, J.-M. Hu, E. Kioupakis, R. Hovden, D. G. Schlom, and

J. T. Heron. Engineering new limits to magnetostriction through metastability in iron-gallium alloys. *Nature communications*, 12(1):2757–2757, 2021.

[62] T. A. Ostler, R. Cuadrado, R. W. Chantrell, A. W. Rushforth, and S. A. Cavill. Strain induced vortex core switching in planar magnetostrictive nanostructures. *Physical Review Letters*, 115(6):067202–067202, 2015.

[63] Qianchang Wang, Xu Li, Cheng-Yen Liang, Anthony Barra, John Domann, Chris Lynch, Abdon Sepulveda, and Greg Carman. Strain-mediated 180° switching in CoFeB and Terfenol-D nanodots with perpendicular magnetic anisotropy. *Applied Physics Letters*, 110(10), 2017.

[64] Cai Chen, Brian D Fu, Marina E Dannecker, Rodrigo U Curiel, Gregory P Carman, and Ab-don E Sepulveda. Exchange stiffness influence on terfenol-d magnetic states. *Multifunctional Materials*, 1(1):014001, oct 2018.

[65] Cheng-Yen Liang, Scott M. Keller, Abdon E. Sepulveda, Wei-Yang Sun, Jizhai Cui, Christo-pher S. Lynch, and Gregory P. Carman. Electrical control of a single magnetoelastic domain structure on a clamped piezoelectric thin film—analysis. *Journal of Applied Physics*, 116(12), 2014.

[66] Francois Cardarelli. *Materials Handbook: A Concise Desktop Reference*. Springer Verlag London Limited, London, 2. aufl. edition, 2008.

[67] MatWeb, LLC. Ruthenium - physical properties. https://apm.matweb.com/reference/elements.aspx, n.d. Accessed: 2025-06-05.

[68] Krishan K Chawla. *Composite Materials: Science and Engineering*. Springer Nature, Cham, fourth edition edition, 2019.

[69] CE Murray and IC Noyan. Finite-size effects in thin-film composites. *Philosophical magazine. A, Physics of condensed matter. Defects and mechanical properties*, 82(16):3087–3117, 2002.

[70] DAVID A. DILLARD. Chapter 1 - fundamentals of stress transfer in bonded systems. In *Adhesion Science and Engineering*, pages 1–44. Elsevier B.V, 2002.

[71] Carl T. Herakovich. *A concise introduction to elastic solids : an overview of the mechanics of elastic materials and structures*. Springer International Publishing, Cham, 1st ed. 2017. edition, 2017.

[72] Zheyuan Yu, Peiran Li, Yin Yao, and Shaohua Chen. An alternative shear lag model for composites with discrete interfaces. *Mechanics of materials*, 176:104530–, 2023.

[73] Maziar Moradi and Siva Sivoththaman. Strain transfer analysis of surface-bonded mems strain sensors. *IEEE Sensors Journal*, 13(2):637–643, 2013.